\documentstyle[prl,aps]{revtex}
\input epsf

\tighten

\begin{document}

%\rline{CWRU-P29-98}

\newcommand{\Bd}{{\dot B}}
\newcommand{\Cd}{{\dot C}}
\newcommand{\fd}{{\dot f}}
\newcommand{\hd}{{\dot h}}
\newcommand{\ep}{\epsilon}
\newcommand{\vp}{\varphi}
\newcommand{\al}{\alpha}
\newcommand{\be}{\begin{equation}}
\newcommand{\ee}{\end{equation}}
\newcommand{\bea}{\begin{eqnarray}}
\newcommand{\eea}{\end{eqnarray}}
\def\gapp{\mathrel{\raise.3ex\hbox{$>$}\mkern-14mu
              \lower0.6ex\hbox{$\sim$}}}
\def\gsim{\gapp}
\def\lapp{\mathrel{\raise.3ex\hbox{$<$}\mkern-14mu
              \lower0.6ex\hbox{$\sim$}}}
\def\lsim{\lapp}
\newcommand{\PSbox}[3]{\mbox{\rule{0in}{#3}\includegraphics{#1}\hspace{#2}}}
\def\Tr{\mathop{\rm Tr}\nolimits}
\def\su#1{{\rm SU}(#1)}

\title{Domain Walls in SU(5)}

\author{
Levon Pogosian
and
Tanmay Vachaspati}
\address
{Department of Physics,
Case Western Reserve University,
10900 Euclid Avenue,
Cleveland, OH 44106-7079, USA.}

\wideabs{
%\twocolumn[
\maketitle

\begin{abstract}
\widetext
We consider the Grand Unified SU(5) model with a small or vanishing
cubic term in the adjoint scalar field in the potential. This gives 
the model an approximate or exact Z$_2$ symmetry whose breaking leads 
to domain walls. The simplest domain wall has the structure of a kink 
across which the Higgs field changes sign ($\Phi \rightarrow -\Phi$) and
inside which the full SU(5) is restored. The kink is shown to be 
perturbatively unstable for all parameters.  We then construct a 
domain wall solution that is lighter than the kink and show it to 
be perturbatively stable for a range of parameters. The symmetry 
in the core of this domain wall is smaller than that outside.  The 
interactions of the domain wall with magnetic monopole is discussed 
and it is shown that magnetic monopoles with certain internal
space orientations relative to the wall pass through the domain wall.
Magnetic monopoles in other relative internal space orientations are 
likely to be swept away on collision with the domain walls, suggesting
a scenario where the domain walls might act like optical polarization
filters, allowing certain monopole ``polarizations'' to pass through
but not others. As SU(5) domain walls will also be formed at small 
values of the cubic coupling, this leads to a very complicated picture
of the evolution of defects after the Grand Unified phase transition.
\end{abstract}
}
%]

\narrowtext

\section{Introduction}
\label{introduction}

Topological defects can be produced at a symmetry breaking
phase transition and would be long-lived relics of the symmetric
phase. If topological defects were produced during a phase 
transition in the very early universe, they could survive 
until the present epoch and thus provide a window to the 
very early universe. The lack of observable defects in the 
present universe helps place strong constraints on particle
physics model building and early universe cosmology.

A prototype symmetry breaking relevant for Grand Unified particle 
physics is 
$$
{\rm SU(5)\rightarrow [SU(3)\times SU(2)\times U(1)]/Z_6} \ . 
$$
The corresponding phase transition
would produce magnetic monopoles. If the only factors 
affecting the evolution of the monopoles are the 
sub-luminal expansion of the universe and monopole-antimonopole 
Coulombic interactions, the monopole abundance grossly violates 
the observed absence of monopoles in the present universe.
The monopole over-abundance problem is solved by invoking 
superluminal universal expansion ({\it i.e.} inflation 
\cite{Gut81}) or by extending the particle physics model so
that the U(1) symmetry gets broken for a short duration
leading to confining forces between monopoles and antimonopoles
\cite{LanPi80}
and thus enhancing their annihilation rate\footnote{There
is another possibility - that the Grand Unified phase transition 
never occurred and hence there never was a monopole
over-abundance problem \cite{DvaMelSen95}.}.
Recently \cite{DvaLiuVac97,PogVac00} we have investigated
the possibility that the Grand Unified phase transition may
also have produced a network of domain walls together with
the magnetic monopoles. These walls would interact with the
monopoles and sweep them away, reducing their abundance to 
an acceptably low level. It is to pursue this scenario in
greater detail that we now study the structure of domain walls 
in the SU(5) model.

The bosonic sector of the SU(5) model is:
\begin{equation}
L = \Tr ( D_\mu \Phi )^2 - V(\Phi )
\label{lagrangian}
\end{equation}
where $\Phi$ is an adjoint of SU(5),
$D_\mu \Phi = \partial_\mu \Phi - ig [X_\mu ,\Phi]$ 
$X_\mu$ are the gauge fields,
and the potential is given by:
\begin{eqnarray}
V(\Phi ) = -m^2 \Tr (\Phi^2) &+& h [\Tr (\Phi^2)]^2 \nonumber \\
   &+& \lambda \Tr (\Phi^4) + \gamma \Tr (\Phi^3) - V_0\ ,
\label{potential}
\end{eqnarray}
where, $V_0$ is a constant that we will choose below.
The SU(5) symmetry is broken to $[SU(3)\times SU(2)\times
U(1)]/Z_6$ if the Higgs acquires a vacuum expectation value
(VEV) equal to
\begin{equation}
\Phi_0 = \frac{\eta}{2\sqrt{15}} {\rm diag}(2,2,2,-3,-3) \ ,
\label{phi0}
\end{equation}
where 
\begin{equation}
\eta = {m \over {\sqrt{\lambda '}}} \ , 
\label{eta}
\end{equation}
\begin{equation}
\lambda ' \equiv h+{7\over {30}}\lambda \ .
\label{lambdaprime}
\end{equation}
For the potential to have its global minimum at $\Phi =\Phi_0$,
the parameters are constrained to satisfy:
\begin{equation}
\lambda \ge 0 \ , \ \ \ \lambda ' \ge 0 \ .
\label{constraint1}
\end{equation}
For the global minimum to have $V(\Phi_0 )=0$, in
eq. (\ref{potential}) we set
\begin{equation}
V_0 = - {{\lambda '}\over 4} \eta^4 \ .
\label{V0}
\end{equation}

The model in eq. (\ref{lagrangian}) does not have any topological 
domain walls because
the vacua related by $\Phi \rightarrow -\Phi$ are not degenerate.
However if $\gamma$ is small, there are walls connecting the
two kinds of vacua that are almost topological. In our analysis
we will set $\gamma =0$, in which case the symmetry of the model
is SU(5)$\times$Z$_2$ and an expectation of $\Phi$ breaks the
Z$_2$ symmetry leading to topological domain walls in addition
to the magnetic monopoles arising from the SU(5) breaking.

In this paper we will study the domain walls present in the 
SU(5)$\times$Z$_2$ model. The simplest kind of domain wall is
the kink that has been studied in a single scalar field model 
with Z$_2\rightarrow 1$ ({\it eg.} \cite{Raj82}). In \cite{PogVac00} 
we studied the interaction of the SU(5) kink with magnetic monopoles 
and found that the monopoles spread out along the kink on collision
and never pass through. This confirmed the conjecture in Ref.
\cite{DvaLiuVac97} that kinks could sweep away magnetic monopoles.
However, the investigations of this paper show that the kink
solution of the SU(5)$\times$Z$_2$ model is unstable to perturbations.
The model contains another domain wall solution that is lighter
than the kink and is perturbatively stable. The adjoint field
does not vanish in the core of these new domain wall solutions 
and hence only a subgroup of the SU(5) is restored at the center.
For this reason, the interactions of these domain walls with magnetic
monopoles is expected to be much more complex (as compared to
the kink), depending on the particular group orientation of the 
monopole relative to the wall.

We will begin our analysis in Sec. \ref{kink} by constructing 
the kink and performing the stability analysis.  Then in Sec. \ref{domainwall} 
we will proceed to construct the domain wall in the model, prove 
that it is lighter than the kink, and show that it is perturbatively
stable for a range of parameters. In Sec. \ref{monodw} we will
consider the interaction of monopoles and domain walls and show
that a monopole whose orientation in group space is aligned with
a colliding domain wall, will pass through and not get swept away.
We further conjecture that monopoles that are misaligned with
the domain wall will be swept away but have not yet shown this.
We draw an analogy of the sweeping out process with that of a
polarization filter that ``sweeps out'' orthogonally polarized 
light and only lets through a certain polarization.

\section{SU(5) kink: solution and stability}
\label{kink}

The kink solution is the Z$_2$ kink along the $\Phi_0$
direction (see eq. (\ref{phi0})). Therefore:
\begin{equation}
\Phi_k =  \tanh ( \sigma z ) \Phi_0 
\label{phik}
\end{equation}
with $\sigma \equiv m/\sqrt{2}$ (see eq. (\ref{lambdaprime})),
and all the gauge fields vanish.
It is straightforward to check that $\Phi_k$ solves the
equations of motion with the boundary conditions 
$\Phi (z=\pm\infty ) = \pm \Phi_0$.
 
As is well-known \cite{Raj82}, the mass (per unit area) of the kink is:
\begin{equation}
M_k = \frac{2\sqrt{2}}{3} {{m^3}\over {\lambda '}} \ .
\label{kinkmass}
\end{equation}

Here we will examine the stability of the kink under general perturbations.
So we write:
\begin{equation}
\Phi = \Phi_k + \Psi 
\label{phiphikpsi}
\end{equation}
Since the kink solution is invariant under translations
and rotations in the $xy-$plane, it is easy to show that the 
perturbations that might cause an instability arise from perturbations
of the scalar field and can only depend on $z$. Therefore we may set
the gauge fields to zero and take $\Psi =\Psi (t,z)$.
%\begin{equation}
%D_x \Phi D_x \Phi + D_y \Phi D_y \Phi + \sum_{i=1}^{3}(
%\frac{1}{2} X_{ix} X_{ix} + \frac{1}{2} X_{iy} X_{iy}) \, \ ,
%\end{equation}
%where $X_{ij}=\partial_i X_j - \partial_j X_i - i g [X_i,X_j]$ are 
%the spatial components of the field strengths.
%This expression is non-negative and is minimized by setting the derivatives
%along the $x-$ and $y-$directions as well as the $x-$
%and $y-$components of the gauge fields to zero. We can also set the
%$z-$components of the gauge fields to zero because the remaining field 
%strengths are vanishing identically.

The Z$_2$ kink is stable and hence we can restrict the
scalar perturbations to be orthogonal to $\Phi_k$. Furthermore,
since the stability of the kink to diagonal perturbations has already
been studied in Ref. \cite{DvaLiuVac97},
we only have to consider perturbations that cannot be diagonalized
by a global SU(3)$\times$SU(2)$\times$U(1) transformation that leaves
the kink invariant.
Therefore we can write:
\begin{equation}
 \Psi  = \sum_{i=1}^{12} \psi^i T^i \, ,
\label{orthogonal}
\end{equation}
where $T^i$ are all generators of SU(5) that do not commute
with $\Phi_0$.

Next we analyze the linearized Schrodinger equation for small excitations
$\psi^i=\psi^i_0 (z) exp(-i\omega t)$ in the background of the kink:
\begin{equation}
[- \partial^2_z - m^2+\phi_k^2(z)(h+\lambda r_i)] \psi^i_0 = 
\omega_i^2 \psi^i_0 \, \ ,
\label{excitations}
\end{equation}
where $\phi_k \equiv \tanh (\sigma z)$ and
$r_i=7/30$. Since this equation is identical for excitations along
any of the $12$ directions, we can drop the index $i$. 
The kink is unstable if there is a solution to 
eq. (\ref{excitations}) with a negative $\omega^2$. 
Substituting eq. (\ref{phik}) into eq. (\ref{excitations}) yields:
\begin{equation}
[-\partial^2_z+m^2(\tanh^2(\sigma z)-1)] \psi_0 = 
\omega^2 \psi_0 \, \ .
\end{equation}
This equation has a bound state solution $\psi_0 \propto $ sech$(\sigma z)$ 
with the eigenvalue $\omega^2=-m^2/2$. Since this result is independent
of the parameters in the potential, we conclude that the kink in SU(5)
is always unstable.

\section{Domain wall}
\label{domainwall}

The domain wall solution is obtained if we choose the 
gauge fields to vanish at infinity and the scalar field
to satisfy the boundary conditions:
\begin{eqnarray} 
\Phi (z=-\infty )= \Phi^- \equiv
  {\eta \over {2\sqrt{15}}}  {\rm diag}(2,-3,2,2,-3) \nonumber \\
         = \eta \sqrt{\frac{5}{12}} (\lambda_3+\tau_3)+
              \frac{\eta}{6}(Y-\sqrt{5}\lambda_8) 
\label{phiminus}
\end{eqnarray}
and
\begin{eqnarray}
\Phi (z=+\infty ) = \Phi^+ \equiv 
  {\eta \over {2\sqrt{15}}} {\rm diag}(3,-2,-2,3,-2) \nonumber \\
         = \eta \sqrt{\frac{5}{12}} (\lambda_3+\tau_3)-
              \frac{\eta}{6}(Y-\sqrt{5}\lambda_8) \ .
\label{phiplus}
\end{eqnarray} 
Here $\lambda_3$, $\lambda_8$, $\tau_3$ and $Y$ are the 
diagonal generators of SU(5):
\begin{equation}
\lambda_3=\frac{1}{2} {\rm diag}(1,-1,0,0,0) \ ,  
\label{lambda3}
\end{equation}
\begin{equation}
\lambda_8=\frac{1}{2\sqrt{3}} {\rm diag}(1,1,-2,0,0) \ ,
\label{lambda8}
\end{equation} 
\begin{equation}
\tau_3=\frac{1}{2} {\rm diag}(0,0,0,1,-1) \ ,  
\label{tau3}
\end{equation}
\begin{equation}
Y=\frac{1}{2\sqrt{15}} {\rm diag}(2,2,2,-3,-3) \ .
\label{ymatrix}
\end{equation}

Note that a local SU(5) transformations can be used to rotate
$\Phi^+$ into $-\Phi^-$ so that the boundary conditions are
like those of the kink with $\Phi(z=+\infty )=-\Phi(z=-\infty )$. 
However, then the solution for
the domain wall will not be diagonal at all $z$. We prefer to
use the above boundary conditions so that the solution is
diagonal throughout.

The domain wall solution can be written as
\begin{equation}
\Phi_{DW}(z)=a(z)\lambda_3+b(z)\lambda_8+c(z)\tau_3+d(z)Y \ .
\label{dwsolution}
\end{equation}
The functions 
$a$, $b$, $c$, and $d$ must satisfy the static equations of motion:
\begin{eqnarray}
a''=[-m^2+(h+\frac{2\lambda}{5})d^2 &+& (h+\frac{\lambda}{2})(a^2+b^2) 
+hc^2 ] a  \nonumber \\
&+&\frac{2\lambda abd}{\sqrt{5}}
\label{aequation}
\end{eqnarray}
\begin{eqnarray}
b''=[-m^2+(h+\frac{2\lambda}{5})d^2 &+& (h+\frac{\lambda}{2})(a^2+b^2) 
+hc^2 ] b \nonumber \\
&+& \frac{\lambda d}{\sqrt{5}}(a^2-b^2) 
\label{bequation}
\end{eqnarray}
\begin{eqnarray}
c''=[-m^2+(h+\frac{9\lambda}{10})d^2 +(h&+&\frac{\lambda}{2})c^2
                  \nonumber \\
&+& h(a^2+b^2)]c
\label{cequation} 
\end{eqnarray}
\begin{eqnarray}
d''=[-m^2 &+& (h+\frac{7\lambda}{30})d^2+(h+\frac{2\lambda}{5})(a^2+b^2) 
          \nonumber\\
        &+&(h+\frac{9\lambda}{10})c^2 ] d + 
        \frac{\lambda b}{\sqrt{5}}(a^2-\frac{b^2}{3}) \ ,
\label{dequation}
\end{eqnarray}
where primes refer to derivatives with respect to $z$.
For reference, the kink solution (eq. (\ref{phik})) 
corresponds to $a(z)=0=b(z)=c(z)$ and $d(z)= \eta \tanh(\sigma z)$.

The equations of motion for $b$ and $c$ and can be solved quite
easily:
\begin{equation}
b(z) = - \sqrt{5} d(z) \ , \ \ \ c(z)=a(z) \ .
\label{bcsoln}
\end{equation}
This is consistent with the boundary conditions in eqs. (\ref{phiminus})
and (\ref{phiplus}). In addition, we require
\begin{equation}
a(z=\pm \infty ) = + \eta \sqrt{5\over {12}} \ , \ \ \ 
d(z=\pm \infty ) = \mp {\eta \over 6} \ .
\label{adbcs}
\end{equation}
Then the remaining equations we need to solve are:
\begin{equation}
a'' = \biggl [-m^2 + \biggl ( 6h +{9\over{10}}\lambda \biggr ) d^2
                   + \biggl ( 2h +{\lambda \over 2} \biggr ) a^2
      \biggr ] a
\label{reducedaeq}
\end{equation}
\begin{equation}
d'' = \biggl [-m^2 + \biggl ( 6h +{{39}\over{10}}\lambda \biggr ) d^2
                   + \biggl ( 2h +{{3\lambda} \over {10}} \biggr ) a^2
      \biggr ] d
\label{reduceddeq}
\end{equation}
These equations can be written in a cleaner form by rescaling:
\begin{equation}
A(z) = \sqrt{{12}\over 5} {a\over \eta} \ , \ \ \
D(z) = - 6 {d\over \eta} \ , \ \ \ 
Z = mz \ .
\label{rescalings}
\end{equation} 
Then
\begin{equation}
A'' = \biggl [ -1 + {{(1-p)}\over 5} D^2 + 
                     {{(4+p)}\over 5}A^2 \biggr ] A
\label{Aeqn}
\end{equation}
\begin{equation}
D'' = \biggl [ -1 + p D^2 + (1-p) A^2 \biggr ] D
\label{Deqn}
\end{equation}
where primes on $A$ and $D$ denote differentiation with respect 
to $Z$, and 
\begin{equation}
p = {1\over 6} \biggl [ 1+ {{5\lambda}\over {12\lambda '}}\biggr ] \ .
\label{pdefn}
\end{equation}
Note that $p\in [1/6,\infty )$ because of the constraints in
eq. (\ref{constraint1}).
The boundary conditions now are:
\begin{equation}
A(z=\pm \infty ) = +1 \ , \ \ \  D(z=\pm \infty ) = \pm 1 \ .
\label{newbcs}
\end{equation}

\begin{figure}[tbp]
\vskip 0.1 truein
\epsfxsize = 0.8 \hsize  \epsfbox{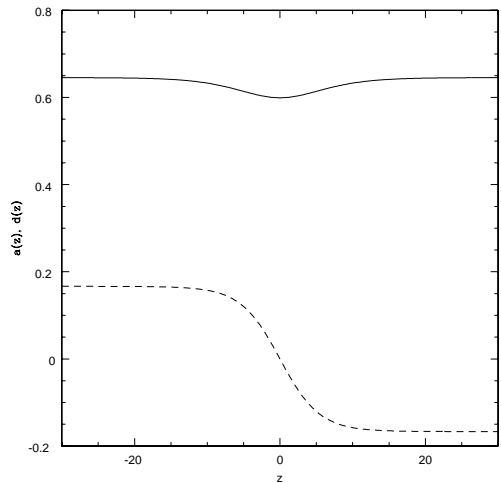}
\caption{\label{danda}
The domain wall solution for $\lambda=1$ and $h=-0.2$ ($p=2.25$).
The solid line shows $a(z)$ and the dashed line shows $d(z)$. 
}
\end{figure}

This system of equations has been solved by numerical relaxation
and a sample solution is shown in Fig. \ref{danda}. 
To find an approximate analytical solution, assume
that $|A''/A| << 1$ is small everywhere. This assumption
will be true for a certain range of the parameter $p$
which we can later determine.
Then the square bracket on the right-hand side of eq. (\ref{Aeqn})
is very small. This gives:
\begin{equation}
A \simeq \biggl [ 
          {{5\over {4+p}}} \biggl \{ 1 - {{(1-p)}\over 5} D^2
               \biggr \} \biggr ]^{1/2}
\label{Aapproxsoln}
\end{equation}
We insert this expression for $A$ in eq. (\ref{Deqn}) and
obtain the kink-type differential equation:
\begin{equation}
D'' = q[-1+D^2]D \ ,
\label{approxDeqn}
\end{equation}
where
\begin{equation}
q = {{6p-1}\over {p+4}} = {{6\lambda}\over {\lambda+60\lambda '}}
\label{qdefn}
\end{equation}
and the solution is:
\begin{equation}
D(Z) \simeq \tanh \biggl ( \sqrt{q\over 2} Z \biggr )
\label{Dapproxsoln}
\end{equation}
The parameter $q$ lies in the interval $[0,6]$. For $q=1$ 
({\it i.e.} $p=1$) it is easy to check that this analytical
solution is exact.

We can now check that our assumption $|A''/A| << 1$ is
self-consistent provided $p$ is not much larger than a few. 

The energy density for the fields $A$ and $D$ can be found
from the Lagrangian in eq. (\ref{lagrangian}) together with
the ansatz in eq. (\ref{dwsolution}), the solution for $b$ and
$c$ in eq. (\ref{bcsoln}) and the rescalings in 
eq. (\ref{rescalings}). The resulting expression for the
energy per unit area of the domain wall is:
\begin{equation}
M_{DW}= {{m^3}\over {12\lambda '}} \int dZ 
         [ 5{A'}^2 + {D'}^2 + V(A,D) ]
\label{tension}
\end{equation}
where,
\begin{eqnarray}
V(A,D) = - 5A^2 &-& D^2 + {{(p+4)}\over 2} A^4 \nonumber \\
          &+& {p\over 2} D^4 + (1-p)A^2 D^2 + 3  \ .
\label{VAD}
\end{eqnarray}
The energy can be found numerically. However, here we will
find an approximate analytic result. We can insert the approximate
solution given above in eq. (\ref{tension}) but this leads to 
an expression that is not transparent. Instead it is more useful 
to consider another approximation for $A$ and $D$:
\begin{equation}
A \simeq 1 \ , \ \ \  
D \simeq \tanh \biggl ( \sqrt{p\over 2} Z \biggr ) \ .
\label{approx2}
\end{equation}
(This approximation is exact for $p=1$.)
A straightforward evaluation then gives:
\begin{equation}
M_{DWapprox} = M_k {{\sqrt{p}} \over 6}
\label{sigmadw}
\end{equation}
where, $M_k$ is given in eq. (\ref{kinkmass}).

We can now compare the domain wall energy to the kink energy. 
If the domain wall is the least energy solution for the 
given boundary conditions, the energy of the exact 
solution for the domain wall will be bounded above by 
the energy of the approximate solution. Note that this 
will be true even if the approximation used to find the 
analytical solution is not good.
Hence this simple argument shows that the domain wall is lighter 
than the kink for $p < 36$, or for $h/\lambda > -6.94/30$. A
numerical analysis shows that the domain wall is lighter than
the kink even in the range $-6.94/30 \ge h/\lambda > -7/30$.
Therefore the domain wall is lighter than the kink for the full
range of parameters specified in eq. (\ref{constraint1}).

Next we study the stability of the domain wall solution. It
is easy to show that the solution is stable to diagonal
perturbations, so here we focus on off-diagonal perturbations.
We write:
\begin{equation}
\Phi=\Phi_{DW}(z)+\sum_{a=1}^{20} \psi^a (z) N^a \, \ ,
\end{equation}
where $N^a$ are the non-diagonal generators of $SU(5)$ and $\psi^a$
are small perturbations satisfying the boundary conditions 
$\psi^a (\pm \infty) = 0$.
Let us first consider the contribution
to the energy density due to fields $\psi^a$.
To second order in perturbations the contributions from different modes,
labeled by index $a$, do not couple. 
%The mode which is contributing the 
%least to the energy density (up to the second order) will be the most 
%likely to become unstable.
A more detailed analysis shows that the mode corresponding to 
\begin{equation}
N_1 \equiv {1\over 2}
                     \pmatrix{0&1&0&0&0\cr
                              1&0&0&0&0\cr
                              0&0&0&0&0\cr
                              0&0&0&0&0\cr
                              0&0&0&0&0\cr} 
\label{N1old}
\end{equation} 
is one of the $8$ modes that are most
unstable. Let $\psi$ be any of these 8 fields. 
The contribution to the energy density due to $\psi$ is:
\begin{eqnarray}
\nonumber
E_{\psi}= \frac{1}{2} (\psi')^2 - 
\frac{m^2}{2} \psi^2 
+ \frac{h}{4}(\psi^2+2a^2+6d^2)^2 \\
+ \frac{\lambda}{4} \psi^2 (a^2+\frac{9}{5}d^2) + 
 {\rm higher} \, {\rm order} \, {\rm terms} 
\, \ ,
\label{epsi}
\end{eqnarray}
where $a$ and $d$ are defined by eq. (\ref{dwsolution}).
As in the case of the kink, we are interested in the
linearized Schrodinger equation for small excitations
$\psi=\psi_0 (z) exp(-i\omega t)$ in the background of
the diagonal domain wall solution. Eq. (\ref{epsi}) leads to
\begin{equation}
[-\partial_z^2 - m^2 + ( 6h +{9\over{10}}\lambda ) d^2
                   +  ( 2h +{\lambda \over 2} ) a^2
      ] \psi_0 = \omega^2 \psi_0 \, \ .
\label{eqforpsi}
\end{equation}
Comparing this with eq. (\ref{reducedaeq}) allows us to write:
\begin{equation}
-\psi_0'' + \frac{a''}{a}\psi_0 = \omega^2 \psi_0 \, \ .
\label{reducedpsieq}
\end{equation}
If $a''/a = 0$, as happens when $p=1$, then there are no non-trivial 
solutions to eq. (\ref{reducedpsieq}) that satisfy the correct 
boundary conditions.
Therefore, the diagonal domain wall solution is stable for at 
least one choice of parameters in the potential, namely, for $p=1$.
By continuity there is a range of parameters
around $p=1$ for which the domain wall is perturbatively stable.

\section{Interaction with monopoles and discussion}
\label{monodw}

To understand the interaction of the domain wall with magnetic
monopoles, it is first useful to understand the structure of
the domain wall core. Since $a(z)$ is non-zero inside the domain wall,
$\Phi_{DW}(z=0) = a(0) (\lambda_3 +\tau_3) \propto {\rm diag}(1,-1,0,1,-1)$.
Therefore, the symmetry inside the core is 
K$~\equiv~$SU(2)$\times$SU(2)$\times$U(1)$\times$U(1). The first SU(2) 
factor arises due to rotations in the 2$\times$2 block with the entries 
equal to 1 in $\Phi_{DW}$ (first and fourth rows and columns). The 
second SU(2) factor is due to the block with entries equal to -1
(second and fifth rows and columns). The two U(1) factors arise since 
there are two diagonal generators of SU(5) aside from those already 
accounted for in the two SU(2) factors, that commute with $\Phi_{DW}(z=0)$.
(We are ignoring any discrete factors that might be present.) The
symmetry group K within the domain wall is to be contrasted
with the full SU(5) symmetry which is restored within the kink. The 
fact that only a subgroup of the SU(5) symmetry is restored in the core
of the wall means that the interaction of the monopole will now depend on 
the particular embedding of the monopole in SU(5) and its orientation 
in internal space relative to the domain wall.

Consider a magnetic monopole whose winding lies in the fourth and fifth
rows and columns of $\Phi$. Staying close to the notation of \cite{PogVac00}
we write the scalar field of such a monopole as:
\begin{equation}
\Phi_M (r) = P(r) \sum_{a=1}^3 x^a \tau_a + M(r) \lambda_8' + N(r) Y \, \ ,
\label{phim}
\end{equation}
where 
$\{\tau_{a}\}$ are the SU(2) generators (see eq. (\ref{tau3})) and
$$
\lambda_8' \equiv 
{1\over{2\sqrt{3}}}{\rm diag}(1,-2,1,0,0)={{\sqrt{3}}\over 2}\lambda_3-
{1\over 2}\lambda_8.
$$
The non-zero gauge fields are: 
\begin{eqnarray}
\nonumber
X_i = \sum_{a=1}^3 X^a_i \tau_a \, \ ,  \\
X^a_i= \epsilon^a_{~ij}
\frac{x^j}{er^2}(1-K(r)) \, .
\label{gaugefields}
\end{eqnarray}
The monopole profile functions, $P(r)$, $M(r)$, $N(r)$ and $K(r)$, 
are solutions of the static equations of motion with boundary
conditions:
\begin{eqnarray}
P(\infty) = \eta \sqrt{5 \over 12} \, \ , 
\ M(\infty) = \eta {\sqrt{5} \over 3} \, \ , \\
N(\infty) = {\eta \over 6} \, \ , \ K(\infty) = 1 .
\end{eqnarray}
When the monopole and the wall are very far from each
other, the combined field configuration can be described by the following 
ansatze:
\begin{eqnarray}
\nonumber
\Phi_{M+DW}= P(r) \frac{c(z')}{c(-\infty)} \sum_{a=1}^3 x^a \tau_a 
       + N(r) \frac{d(z')}{d(-\infty)} Y \\
    + M(r) \left[ \frac{\sqrt{3}}{2}\frac{a(z')}{a(-\infty)}\lambda_3
     -\frac{1}{2}\frac{b(z')}{b(-\infty)}\lambda_8 \right]
\, \ ,
\label{mpdw} 
\end{eqnarray}
where $z'=z-z_0$ and $z_0$ is the initial position of the domain wall.
When $r$ is small $\Phi_{M+DW} \rightarrow \Phi_M$ (eq. (\ref{phim})) and
when $z'$ is small $\Phi_{M+DW} \rightarrow \Phi_{DW}$ (eq. (\ref{dwsolution})) 
along the $z$-direction. 
The gauge fields are the same as for the monopole alone.
We have purposely chosen the embedding of the monopole so
that all interesting dynamics of the monopole-wall interaction
is restricted to the fourth and fifth rows and columns
of $\Phi_{M+DW}$. This follows from the equations of motion and the
commutation properties of the generators appearing in the ansatze (\ref{mpdw}).
Let us then only consider the relevant part of the $\Phi_{M+DW}$ matrix:
\begin{eqnarray}
\nonumber
\Phi_{2\times 2} \equiv {1\over 2} P(r) \frac{a(z')}{a(-\infty)}
               \pmatrix{z&x-iy\cr
                   x+iy&-z} \, \\
        - {3\over 2\sqrt{15}} N(r) \frac{d(z')}{d(-\infty)}
        \pmatrix{1&0\cr
                 0&1\cr} \, . 
\label{relevantpart}
\end{eqnarray}
The form of $\Phi_{2\times 2}$ suggests that the 
only field that is going to be considerably affected by the domain wall 
is $N(r)$ because $a$ is roughly constant in space.
There is no angular dependence in the term with $N(r)$ in eq. (\ref{phim})
and hence $N(r)$ does not contribute to the winding of the monopole. 
Therefore, we do not expect the wall to affect the winding.
Essentially the reason is that the SU(2) subgroup in which the
monopole winding is located is not restored on the wall.
We have checked that the monopole passes right through the wall
explicitly in this case by numerically colliding the monopole and 
the wall.

If the magnetic monopole winding lies in the first and fourth block,
it will experience a region of restored SU(2) symmetry inside the
domain wall and hence we conjecture that such monopoles will unwind
on the wall. If our conjecture is correct, the domain walls behave 
similarly to optical polarization filters, allowing monopoles with 
certain internal space polarizations to pass through and annihilating 
other polarizations. The detailed analysis of all possible monopole 
embeddings is a challenging project, both numerically and analytically, 
since one cannot avoid dealing with a large number of the fields 
present in SU(5). 

There is another possibility that is worth pointing out. If a
domain wall and a magnetic monopole are misaligned in internal
space, it may not be possible to superpose the two solutions
so as to get a monopole and a domain wall together. (Such a
situation is known to occur when attempting to construct
multimonopole or multistring solutions.) Then it is likely that 
there will be a long range force between the domain wall and 
misaligned monopole that will bring them together. On coming together
the monopole could get annihilated on the wall, or else in
some cases, it may get aligned and then pass through the wall.

Our considerations point to a very complicated aftermath of the
GUT phase transition. Domain walls and magnetic monopoles would
both be produced and would start interacting. The outcome of an
interaction would depend on the internal space orientations of
the monopole relative to the domain wall. Any given domain wall 
would be transparent to some monopoles but not to others. The
relaxation of the system would depend on whether a monopole
encounters a sufficient number of randomly oriented (in internal
space) domain walls, at least one of which might sweep it away.
It remains to be seen if domain walls can provide a means to
solve the cosmological monopole over-abundance problem.

\acknowledgements

We would like to thank Mark Trodden for many useful discussion.
This work was supported by the DOE.

\end{document}